\documentclass[moor]{informs1}

\usepackage{natbib}
 \NatBibNumeric
 \def\bibfont{\small}%
 \bibpunct[, ]{[}{]}{,}{n}{}{,}%

\usepackage[colorlinks=true,breaklinks=true,bookmarks=true,urlcolor=blue,
     citecolor=blue,linkcolor=blue,bookmarksopen=false,draft=false]{hyperref}

\def\EMAIL#1{\href{mailto:#1}{#1}}

\TheoremsNumberedThrough     

\EquationsNumberedThrough    
\usepackage{booktabs}
\usepackage{booktabs}
\usepackage{tikz}
\usetikzlibrary{matrix}
\usetikzlibrary{arrows,automata}
\usepackage{amsmath,amssymb,url}
\usepackage{algorithm,algpseudocode}
\usepackage{graphics}
\usepackage{float}
\usepackage[T1]{fontenc}
\usepackage[utf8]{inputenc}
\usepackage{subcaption}
\usepackage{algpseudocode}
\usepackage{tabularx}
\usepackage{booktabs}
\usepackage{enumerate}
\usepackage{verbatim}
\usepackage{pgf}
\usepackage{theorem}

\usepackage{relsize}
\usepackage{exscale}

\usepackage{setspace}
\usepackage{color}
\definecolor{light-gray}{gray}{0.95}
\def\red#1{{\color{red}#1}}
\def\blue#1{{\color{blue}#1}}
\long\def\old#1{}

\def \cut{\text{c}}

\def \P{\mathbb{P}_{I_0}^\rho}
\def \E{\mathbb{E}_{I_0}^\rho}
\def \hI{\hat{I}}

\long\def\old#1{}

\def\red#1{{\color{red}#1}}
\def\del#1{\blue{[\sout{#1}]}}

 \RUNAUTHOR{ Drakopoulos,  Ozdaglar, and  Tsitsiklis}

 \RUNTITLE{Efficient Curing}

\TITLE{An efficient curing policy for epidemics on graphs\footnote{Research partially supported by {the Draper Laboratories and NSF grant CMMI-1234062. A preliminary version of some of the results in this paper are included in a paper submitted to the 53rd IEEE Conference on Decision and Control, December 2014.}}} 

\ARTICLEAUTHORS{%
\AUTHOR{Kimon Drakopoulos, Asuman Ozdaglar, John N. Tsitsiklis\\ }
\AFF{ Laboratory of Information and Decision Systems, Massachusetts Institute of Technology, Cambridge, MA 02139, \EMAIL{\{kimondr,asuman,jnt\}@mit.edu}.\\ \vspace{1cm}\large{\today}}
}

\ABSTRACT{We  provide a  dynamic policy for the rapid containment of a {contagion} process modeled as an SIS epidemic on a bounded degree {undirected} graph with $n$ nodes.
{We show that if the budget $r$ of curing resources available at each time is 
$\Omega(W)$, where $W$ is the CutWidth of the graph, and also of order $\Omega(\log n)$, then the expected time until the extinction of the epidemic is of order $O(n/r)$, which is within a constant factor from optimal, as well  as sublinear in the number of nodes. Furthermore, if the CutWidth increases only sublinearly with $n$, a sublinear expected time to extinction is possible with a sublinearly increasing budget $r$.}
}
\begin{document}
\maketitle

\section{Introduction}
Many {contagion} processes over large networks can lead to  costly cascades unless controlled by outside intervention. Examples include epidemics spreading over a population of individuals, {viruses} attacking a network of connected computers, or financial contagion {in} a network of banks. In this paper we study how this type of contagion can be prevented or contained by dynamically curing some of the infected nodes under  a budget constraint on the amount of curing resources that can be deployed at {each} time.

More specifically, we consider a canonical SIS epidemic model on an {undirected} graph\footnote{{Our results actually are easily generalized to the case of directed graphs.}} with $n$ nodes, with a common infection rate {along any {edge} that connects an infected and a healthy node}, 
and node-specific curing {rates} $\rho_{v}(t)$ {at each node ${v}$. The curing rates are to be chosen according to a curing policy which is based on the past history of the process and the network structure, subject to an upper bound on the total curing rate $\sum_{{v}} \rho_{{v}}(t)$.} 

{In a companion paper,} \cite{DOT14} we characterize the {cases} for which a contagion process can be rapidly contained, i.e., the expected time to extinction {can be made} sublinear in the number {$n$} of nodes using a sublinear curing budget. {Our characterization involves the {CutWidth} of the underlying graph.  Intuitively, the CutWidth measures the {required budget of} curing resources {in} a {simpler deterministic} curing problem, in which infected nodes are cured one at a time, subject to the constraint that the {number of edges} between healthy and infected nodes is at all times less than or equal to the {budget} of curing resources. In \cite{DOT14}, 
{we establish that if the  CutWidth increases at least linearly with {$n$}, then a sublinear {(in $n$)} expected time to extinction is impossible {with a sublinear budget $r$.}}
{On the other hand, \cite{DOT14} provides} a {\it nonconstructive proof} that for {graphs} with sublinear CutWidth and bounded degree, there exists a dynamic policy that achieves sublinear expected  time to extinction using only a sublinear budget,}
{for any set of initially infected nodes.} 
The main contribution of the present paper is the {\it construction of a specific}  policy with {the latter} desirable properties.

{Our} policy is based on {a  combinatorial result which states the following. Given an initial set of infected nodes, nodes can be removed from that set, one at a time, in way that  
the maximum cut {(number of edges)} between healthy and infected nodes encountered during this process is upper bounded by the {{sum} of the} CutWidth of the graph and the cut {associated with the initial set. Let us refer to the sequence of subsets encountered during this process as a {\it target path}.} The main idea {underlying our} policy is to allocate {the entire curing} budget to  appropriate nodes {so that we stay most of the time, with high probability,}  on {or near} the target path. 
We show that this is indeed possible,  {as long as the curing resource budget scales in proportion to the CutWidth.}
We also
show that the policy is optimal {(within a multiplicative constant) if} the available budget is also {$\Omega(\log n)$}.\footnote{We write $f(n){=}o(g(n))$ {if $\lim_{n\to\infty} f(n)/g(n)=0$.}
We write $f(n){=}\Omega(g(n))$ if {$\liminf_{n\to\infty} f(n)/g(n)>0$.}
Finally, we write $f(n){=}O(g(n))$ if {$\limsup_{n\to\infty} f(n)/g(n)<\infty$.}} 

A similar {model, but in which the curing rate allocation is done statically (open-loop)}
 has been studied in \cite{Cohen,gourdin, chung,preciado}, 
but the proposed methods {were} either heuristic or based on  mean-field approximations of the evolution process. {Closer} to our work, {the authors of  \cite{Borgs10} let the curing rates be proportional} to the degree of each node --- independent of the current state of the {network, which means that curing resources may be wasted on healthy nodes.} On a graph with bounded degree, {the policy in \cite{Borgs10}} achieves sublinear  time to extinction, {but requires a} curing budget that is proportional to the number of nodes. In contrast, our policy achieves the same performance (sublinear  time to extinction) for all bounded degree graphs with small CutWidth by properly focusing the curing resources. As an extreme example, consider a  line graph with $n$ nodes, and assume that the $n/2$ leftmost nodes are {initially} infected. The degree-based policy of \cite{Borgs10} requires a total budget proportional  to {$n$} and allocates it proportional to the degree. {In contrast, our policy
can achieve sublinear expected  time to extinction with a {$\Omega(\log n)$ but} sublinear budget. This is because, instead of allocating the available budget to all nodes, our policy focuses on specific nodes on the boundary between healthy and infected nodes, in this instance on the rightmost infected node.} By extending this idea, our policy  achieves {a similar} improvement for all graphs with sublinear CutWidth.

The rest of the paper is organized as follows. {In Section \ref{model} we present the details of the model that we are considering. In Section \ref{sec:combinatorics} we introduce the CutWidth and establish the combinatorial result mentioned earlier.} In Section \ref{sec:CuRe} we present the policy and analyze its performance.
{In Section \ref{s:cor} we develop some corollaries that demonstrate the possibility of fast extinction using a sublinear budget and the approximate optimality of our policy in a certain regime. We also mention some examples.}
Finally, in Section
\ref{concl} we offer some closing remarks.

\section{The Model} \label{model}
We consider a {network}, represented by a {connected} {undirected} graph  $G=(V,E)$, where $V$ denotes the set of nodes and  $E$ denotes the set of edges. {We use $n$ to denote the number of nodes.} Two nodes $u,v \in V$ are {\it neighbors} if $(u,v) \in E$. We denote by $\Delta$ the maximum {of the node degrees.} 

We assume that {the} nodes in a set $I_0{\subseteq V}$ are initially infected and {that} the infection spreads according to a controlled contact process where the rate at which infected {nodes} get cured is  determined by a network controller. Specifically, each {node} can be in one of two states: {\it infected}  or {\it healthy}. The controlled contact process --- {also {known} as the SIS epidemic {model}} --- on $G$ is a 
{right-continuous}, continuous-time Markov process 
{$\{I_t\}_{t\geq 0}$ on the state space $\{0,1\}^V$, where $I_t$ stands for the set of infected nodes at time~$t$.} {We refer to $I_t$ as the {\it infection process}.}

State transitions at each node occur independently according to the following dynamics.
\begin{enumerate}[a)] \item {The process 
is initialized at the given initial state $I_0$.}
\item {If a node $v$ is healthy, i.e., if $v\notin I_t$, the transition rate associated with a change of the state of that node to being infected is  equal to an infection rate $\beta$ times the number of infected neighbors of $v$, that is,
$$\beta \cdot  \big|\{(u,v)\in E: u\in I_t\}\big|,$$
where we use $|\cdot|$ to denote the cardinality of a set.}
{By rescaling time, we can and will assume throughout the paper that $\beta=1$.} 

\item {If a node $v$ is infected, i.e., if $v\in I_t$, the transition rate associated with a change of the state of that node to being healthy is equal to a curing} rate  
 $\rho_v(t)$ that is determined by {the} network controller, as a function of  the current and past states of the {process.} We are assuming here that the network controller has access to {the entire} past evolution of the process.
\end{enumerate}

We assume a {\it budget constraint} 
{of the form}    
\begin{equation}
\sum_{v \in V} \rho_v(t)\leq r,  \label{eq:budgetconstr}
\end{equation}
{for each} time instant $t$, reflecting the fact that curing is costly. 
 A {\it curing policy}  is a  mapping {which at any time $t$ maps the past history of the process to a curing vector $\rho(t)=\{\rho_v(t)\}_{v \in V}$} that satisfies  (\ref{eq:budgetconstr}).

\old{We assume that {the process} {$I_t$} is right-continuous.
Let  $\mathbb{P}^\rho_{I_0}$ be the  probability measure induced by ${I_0}$ and the policy $\rho$. The expectation corresponding to $\P$ is denoted by $\E$. }

We define the {\it  time to extinction} as the {time until the process reaches the absorbing state where all nodes are healthy:} 
$$\tau = {\min}\{t\geq0 : {I_t=\emptyset}\}. 
$$
{The {\it expected  time to extinction} {(the expected value of $\tau$)  is the performance measure that we will be focusing on.}

\section{Graph theoretic  {preliminaries}}\label{sec:combinatorics}
In this section we introduce {the notions of a cut and of the CutWidth}  that will be used {in} the description of {our}  policy. 
We state some of their properties and then proceed to develop}
a key combinatorial result {that}  
{will play a critical role in the analysis of our policy's performance.} 
Throughout, we assume that we are dealing with a particular given graph $G$.

\subsection{CutWidth}\label{s:cw}
{For convenience, we will be using the shorthand term ``bag'' to refer to `` a subset of $V$.'' We also use the following notation.
For any two bags $A$ and $B$, and any $v\in V$, we let}
\[
A \setminus B= \{v \in A: v \notin B\},
\]
and
\[
A-v=A \setminus \{v\}.
\] 
{We also use $A^c$ to denote the complement, $V\setminus A$ of $A$.}

We next define the concept of a {\it monotone crusade}. A monotone crusade from {a bag} $A$ to {another bag} $B$, where $B \subseteq A$, is a {finite} sequence of bags that starts {with} $A$ and {ends with} $B$, {so} that at {each} step of {the} sequence {no nodes are added   (cf.\ Part (iii) of Definition \ref{d:crus}), and 
exactly one node  is removed}
(cf.\ Part {(iv)} of Definition {d:crus}).

\begin{definition}\label{d:crus}
For any two bags $A$ and $B$, {with $B \subseteq A$, a (monotone) crusade from $A$ to $B$, or  ($A \downarrow B$)-\textbf{crusade} for short},  is a sequence ${\omega=}(\omega_0,\omega_1,\ldots,\omega_{{k}})$ of bags of length $|\omega|=k{+1}$, with the following properties:
\begin{enumerate}[(i)]
\item $\omega_0 = A$,
\item $\omega_{k}=B$, 
\item {$\omega_{i+1}\subset \omega_i$, for  $i=0,1,\ldots,k-1$, and}
\item $|\omega_i \setminus \omega_{i+1} |= 1$, for  {$i=0,1,\ldots,k-{1}$. }
\end{enumerate}
\end{definition}
We denote by ${\mathcal{C}}(A \downarrow B)$  the set of all $(A \downarrow B)$-crusades. 

The number of edges connecting {a} bag {$A$} with its complement is called the cut of the bag. {Its importance lies in that it is equal to the total rate at which new infections occur, when the set of currently infected nodes is $A$.}
\begin{definition} {For any bag $A$, its \textbf{cut}, $\cut(A)$, is defined as the cardinality of the set of edges
$$\big\{(u,v): u\in A,\ v\in A^c\big\}.$$}
\end{definition}

\noindent 
{In Proposition \ref{prop:cutproperties} below, we record, without proof, an elementary property of cuts.}

\begin{proposition}
\label{prop:cutproperties}
{For any two bags $A$ and $B$, we have 
$$\cut(A\cup B)\leq \cut(A)+\cut(B)\leq \cut(A)+\Delta\cdot |B|.$$}
\end{proposition}

We define the width of a {monotone} crusade $\omega$ as the maximum cut encountered during the crusade. {Intuitively, this is the largest infection rate to be encountered if the nodes were to be cured according to the sequence prescribed by the crusade {deterministically, if no {new} infections {happen}  in between.}}

\begin{definition}
Given an ($A \downarrow B$)-crusade $\omega=(\omega_0, \ldots, \omega_{{k}})$, its \textbf{width} $z(\omega)$ is defined by
\[
z(\omega)=\max_{ {0} \leq i \leq {k}}\{\cut(\omega_i)\}.
\]
\end{definition}

{We next define what we call the impedance of a bag $A$, as the minimum possible width among the $\mathcal{C}(A\downarrow \emptyset)$-crusades. This minimization captures the objective of finding a crusade along which the total infection rate is always small.}

\begin{definition} The \textbf{impedance} {$\delta(A)$} of {a bag $A$} {is defined by}    
\begin{equation}
\delta(A)\doteq \min_{\omega \in \mathcal{C}(A\downarrow \emptyset)} z(\omega). \label{eq:impedancedef}
\end{equation}
{For the special case where $A=V$, the impedance is known as the {\bf CutWidth} \cite{Mak89,BiSe91, Chu85,Kor93}, and will be denoted by $W$.}
\label{def:impedance}
\end{definition}

{We say that a (monotone) crusade ($A \downarrow B$)-crusade $\omega=(\omega_0, \ldots, \omega_{k-1})$ is {\it optimal} if it attains the minimum in Eq.\ 
\eqref{eq:impedancedef}.}
It can be seen that the {impedances  satisfy}  the Bellman equation: 
\begin{equation} \label{eq:bellman}
{\delta(A) = 
\max\big\{ \cut(A),\ \min\{\delta(B): B\subseteq A,\  |A {\setminus}  B|= 1\}
\big\}.}
\end{equation}
{Furthermore, along an optimal crusade, we have $\delta(\omega_{i+1})\leq \delta(\omega_i)$, for $i=0,1,\ldots,k-1$. Finally, we note that $\cut(A)\leq \delta(A)$.}

\subsection{Impedance and CutWidth}
In this {subsection} we discuss 
{the relation between}
the impedance of {an arbitrary} bag and the CutWidth. 
{The impedance of a bag $A$ is at least $\cut(A)$, which {in general} may be much larger than the CutWidth.}\footnote{{As an example, consider a line graph, and let $A$ be the set of even-numbered nodes. Then, $\cut(A)$ is approximately $n$, whereas the CutWidth of the line graph is equal to 1.}}
{This is a concern because the stochastic nature of the infections can always bring the process to a bag with high impedance, and therefore high subsequent infection rates.  The next lemma provides an upper bound on the impedance of a bag $A$ in terms of the CutWidth $W$ of the graph and the cut of $A$.   Its proof is given in the Appendix.}

\old{\begin{figure}
\centering
\includegraphics[width=0.4\textwidth]{lineexample}
\caption{Consider a line graph with $n$ nodes and a bag that contains all even numbered nodes.   For any monotone crusade, the bag obtained after the first step, that is, after curing one node, has a cut that is proportional to $n$, and therefore the impedance of the bag is proportional to $n$. The impedance of the whole graph is $1$ and a monotone crusade which achieves the minimal width  cures nodes one by one starting from the rightmost node. Therefore, the impedance of the bag is much higher than the impedance of the graph.
\red{[CAN YOU MAKE THE CAPTION TO BE RIGHT-JUSTIFIED RATHER THAN CENTERED TEXT?]}
}
 \label{fig:nonmonotonicity}
\end{figure}
}

\begin{lemma}\label{lem:monotonewithcut}
For any   bag $A$, {we have}
\[
\delta(A) \leq {W}+\cut(A).
\]
\end{lemma}

\section{The CURE policy }\label{sec:CuRe}

In this section, we present our curing policy and we study the {resulting}  expected  time to extinction, {starting from an arbitrary initial set of infected modes.} {Loosely speaking, the policy, at any time, tries to follow a certain desirable (monotone) crusade, called a target path, by allocating all of the curing resources to a single node, namely, the node that should be removed in order to obtain the next bag along the target path. On the other hand, this ideal scenario {may be} interrupted by  infections, at which point the policy shifts its attention to newly infected nodes, and attempts to return to a bag on the target path. It turns out that under certain assumptions, this is successful with high probability and does not take too much time. However, with small probability, the process veers far off from the target path; in that case the policy ``restarts'' in a manner that we will make precise in the sequel.}

{It is quite intuitive (and formally established in \cite{DOT14}) that a fast (sublinear)  time to extinction may not be possible if the curing budget is smaller than the CutWidth. For this reason, we focus on the regime where the curing rate is at least proportional to the CutWidth, and more concretely, on the regime where $r\geq 4 W$, which we henceforth assume.}

{Under the above assumptions on the budget $r$, and the additional assumptions that $r=\Omega(\log n)$ {and $r\geq 8 \Delta$}, we will construct a policy whose expected  time to extinction is $O(n/r)$; cf.\ Theorem \ref{thrm:upperbound} and Corollary~\ref{coro}.}

\subsection{Description of the CuRe policy}\label{s:descr}

\old{\begin{figure*}
\includegraphics[width=\textwidth]{policydiag}
\caption{An attempt is a sequence of segments. Each segment starts with a path-following phase. When an infection occurs an excursion starts. If the excursion is short the process returns to a path-following phase. If the excursion is long a waiting period starts. When the waiting period is over a new attempt begins.}\label{fig:attemptdiagram}
\end{figure*}
\begin{figure}
\centering
\includegraphics[width=0.45\textwidth]{blockdiagram}
\caption{A flowchart of the policy.}\label{fig:block}
\end{figure}}

\old{
We start with some notation. We let $T_0=0$ and, for $i\geq 1$, let $T_i$ be the $i$th time at which the process $\{X_v(t)\}$ makes a transition.
Let $i^*$ be the total number of transitions. In particular, if the process eventually reaches the absorbing state where all nodes are cured, then $i^*<\infty$.
Note that the set of infected nodes $I_t$ takes a constant value during the interval
$[T_i,T_{i+1})$, where $0\leq i\leq i^*$. We will use $\hI_i$ to denote that constant value.}}

The CURE policy is defined  {hierarchically:}  it consists of a {sequence} of attempts;  each attempt consists of a {waiting period, followed by} a sequence of segments; finally, each segment consists of {a path-following phase and an excursion. Each attempt, at the end of its waiting period, determines a path that its segments will try to follow. If an attempt fails to cure all nodes, a new attempt is initiated.}
{We now proceed to describe in detail  a typical attempt. 
\old{See  Figure \ref{fig:attemptdiagram} for an illustration of the different segments and phases, and Figure \ref{fig:block} for an associated flowchart.}}

\noindent  {{ \bf Waiting period.} A typical attempt starts at some bag $A$, with a waiting period. (If this is the first attempt, then $A=I_0$. Otherwise, it is the bag at the end of the preceding attempt.) During the waiting period, all curing rates $\rho_v(t)$ are kept at zero. The waiting period ends  at the first subsequent time
that\footnote{Note that the waiting period is guaranteed to terminate in finite time, with probability 1. This is because if it were infinite, then healthy nodes would keep getting infected until eventually $I_t=V$. But $\cut(V)=0$, which means that at some point the condition  $\cut(I_t)\leq{r}/{8}$ would be satisfied and the waiting period would be finite, a contradiction.} 
$$\cut(I_t)\leq{r}/{8}.$$
Let $B$ be the bag $I_{t}$ right at the end of the waiting period, and
let   $\omega^B =(\omega^B_0,\ldots,\omega^B_{|B|})$ the corresponding optimal crusade, which we refer to as the {\it target path}.}

{
\noindent{ \bf Segments.} 
Each segment of an attempt starts either at the end of  the waiting period or at the end of a preceding segment of the same attempt. 
In all cases, the segment starts with a bag on the target path. For the first segment, this is guaranteed by the definition of the target path. For subsequent segments, it will be guaranteed by our specifications of what happens at the end of the preceding segment.
}

{\noindent
{\bf Path-following phase.} 
Let $v_1,\ldots,v_m$ be the nodes in the bag at the beginning of a segment, arranged in the order according to which they are to be removed along the target path. For example, the bag at the beginning of the segment is $\{v_1,\ldots,v_m\}$, the next bag is $\{v_2,\ldots,v_m\}$, etc. 
During the path-following phase, the entire curing budget is first allocated to node
$v_1$ until it gets cured, then to node $v_2$, etc.
The phase ends when either: 
\begin{itemize}
\item[(i)] all nodes have been cured, i.e., $I_t=\emptyset$; in this case, the attempt is considered successful and the process is over.
\item[(ii)] an infection occurs; in this case, $I_t$ is outside the target path, and the segment continues with an excursion phase.
\end{itemize}
}

{\noindent
{\bf Excursion.} If a path-following phase ended with an infection,  the  {policy enters an excursion phase}. Let $C$ be the next bag (on the target path), namely the bag that would have been reached if a node was cured before the infection happened. 
We define  $D_t=I_t \setminus C$; this is the set of infected nodes that do not belong to the next bag on the target path. The goal during the excursion is to cure the nodes in $D_t$, so as to reach the bag $C$ on the target path, {and {then} start a } subsequent segment. With this purpose in mind,   during an excursion, we allocate all the available budget on an arbitrary node in $D_t$. The excursion ends when $|D_t|$ becomes either zero  or at least $r/8 \Delta$. 
\begin{itemize}
\item[(i)]
If the excursion ends with $|D_t|=0$, we say that we have a 
{\it short} excursion. At that time, we are back on the target path, with $I_t=C$, and we are ready to start with the next segment.
\item[(ii)]
If on the other hand $|D_{t}|\geq r/8 \Delta$, we say that the excursion was {\it long}, and that the attempt has failed. In this case, the attempt has no more segments, and a new attempt will be initiated, starting with a waiting period.
\end{itemize}}

\subsection{Performance analysis --- Outline}
{We now proceed to establish an upper bound on the  expected time to extinction, under the assumption that $r\geq 4 W$,  for any set of initially infected nodes.} 
If the process always stayed on the target path, that is if we had no infections,   the expected  time to extinction would be the time until {all nodes (at most $n$ of them) were cured.
Given that nodes are cured at a rate of $r$,
 the expected  time to extinction would have been  $O(n/r)$. On the other hand, infections do delay the curing process, by initiating excursions, and we need to show that these do not have a major impact.}
 
There are two kinds of excursions to  consider,  {\it short} {ones, at the end of which}  $|D_t|=0$, and {\it long} {ones, at the end of which} $|D_t|{\geq}r/8 \Delta$. 
During {an} excursion, the size of $D_t$ {(the ``distance'' from the target path) is} at most $r/8 \Delta$. {Using also an upper bound on the size of the cut along the target path, we can show that the infection rate during an excursion is smaller than the curing rate. For this reason, during an excursion, the process $|D_t|$ has a downward drift. As a consequence, using a standard argument, the expected duration of an excursion is small and there is high probability that the excursion ends with $|D_t|=0$, so that the excursion is short and we continue with the next segment.} As a result, the expected {duration of an attempt} behaves similar to the case {of no excursions and is also of order $O(n/r)$. 
Finally, by viewing each attempt as an independent trial, we can establish an upper bound for the overall policy.}
A formal version of this argument is the content of the rest of this section.

\subsection{Excursion analysis} 

Let us focus on a particular excursion, and let, for simplicity, $M_t=|D_t|$. 
The process $M_t$ evolves on the finite set $\{0,1,\ldots,r/8\Delta\}$.
(For simplicity, and without loss of generality, we assume that $r/8\Delta$ is an integer.)
Recall that $C$ was defined as the bag on the target path that we were trying to reach at the end of the excursion. The difference $D_t$ at the time that the excursion starts consists of exactly two nodes: the node that we were trying to remove just before the excursion started and the node outside the target path that got infected. Thus, the process $M_t$ is initialized at 2, at the beginning of the excursion. The process $M_t$ 
is stopped as soon one of the two boundary points, 0 or $r/8\Delta$, is reached.
At each time before the process is stopped, there is a rate equal to $r$ of downward transitions. Furthermore, there is a rate $\cut(I_t)$ of upward transitions, corresponding to new infections.

\begin{lemma}\label{l:tr}
The rate $\cut(I_t)$ of upward transitions during an excursion satisfies $\cut(I_t)\leq r/2$.
\end{lemma}
\proof{Proof:}
The definition $D_t=I_t\setminus C$  implies that $I_t\subseteq C \cup D_t$. 
Consequently, 
\begin{align}\label{eq:bb}
\cut(I_t) &\leq \cut (C) + \cut(D_t) \leq \cut(C) +\Delta\cdot |D_t| \\  \nonumber
&= \cut(C)+\Delta\cdot M_t\leq \cut(C)+\frac{r}{8}.
\end{align}
We have used here Proposition \ref{prop:cutproperties}, in the first and second inequality,  together with the fact $M_t\leq r/8\Delta$.

On the other hand, $C$ is on the target path {associated with $B$, the bag obtained at the end of the waiting period.}  As remarked at the end of Section \ref{s:cw}, the impedance does not increase along an optimal crusade, and therefore, $\delta(C)\leq \delta(B)$. 
 Using also Lemma \ref{lem:monotonewithcut}, we have
 $$\delta(C)\leq \delta(B) 
\leq 
 W+\cut(B).$$
Recall now that a waiting period ends with a bag whose cut is at most $r/8$. Therefore,
$\cut(B)\leq r/8$. It follows that $\cut(C)\leq W+r/8$. Using this fact, together with the assumption $r\geq 4W$ and Eq.\ \eqref{eq:bb}, we obtain
$$\cut(I_t)\leq \cut(C)+\frac{r}{8} \leq \Big(W+\frac{r}{8}\Big)+ \frac{r}{8}
\leq \frac{r}{4}+\frac{r}{8}+\frac{r}{8}=\frac{r}{2}.$$
\endproof

We now establish the properties of the excursions that we have claimed; namely, that excursions are short, with high probability, and do not last too long.

\begin{lemma}\label{lem:excursionproperties}
\begin{itemize}
\item[a)] The probability that the excursion is long is at most
$$p=\frac{2^2-1}{2^{r/8\Delta} -1}.$$
\item[b)] The expected length of an excursion is upper bounded by $4/r$.
\end{itemize}
\end{lemma}
\proof{Proof:}
\begin{itemize}
\item[a)] 
Using Lemma \ref{l:tr}, the process $M_t$ is stochastically  dominated by a process $N_t$ on the same space $\{0,1,\ldots,r/8\Delta\}$,  which is initialized to be equal to the value of $M_t$ at the beginning of the excursion (which is 2), has a rate $r$ of downward transitions, a rate $r/2$ of upward transitions, and stops at the first time that it reaches one of the two boundary values. Note that the ratio of the downward to the upward drift is equal to 2.
The probability, denoted by $p$, that the process $N_t$ will first reach the upper boundary is a well-studied quantity and is given by the expression in part (a) of the lemma.
The proof is standard and can be found in Section 2.1 of \cite{Peres} (for a non-martingale based proof) or Section 2.3 of \cite{Stee00} (for a martingale based proof). Since $M_t$ is stochastically dominated by $N_t$, the probability that $M_t$ will first reach the upper boundary is no larger.

\item[b)] 
For simplicity, let us suppose that the excursion starts at time $t=0$. We define the process 
\[
H_t=M_t + \frac{r}{2}t
\]
and the stopped version, ${\hat H}_t$ which stops at the time $T$ that the excursion ends.
It is straightforward to verify that $\hat{H}_t$ is a supermartingale, because the upward drift of the process is $\beta \cut(I_t) \leq r/2$ and the downward drift is $r$, so that the total downward drift at least $r/2$. Furthermore, $H_0=M_0=2$.
Using Doob's optional stopping theorem we obtain
$$
2=\mathbb{E}[M_0]= \mathbb{E}[H_0] \geq\mathbb{E}[H_T]+\frac{r}{2}\cdot T \geq \frac{r}{2}\cdot T,
$$
from which we conclude that 
\[
\mathbb{E}[T] \leq \frac{4}{r}.
\]
\end{itemize}
\endproof

Note that if $r\geq \alpha \log n$, where $\alpha$ is a sufficiently large constant, 
then $p$ can be made smaller that $1/n^2$, so that $np$ tends to zero. We will be using this observation later on.
We will now bound the length of a waiting period.

\begin{lemma} \label{lem:detour}
The expected length of a waiting period is bounded above by $8n/r$.
\end{lemma}
\proof{Proof:}
A waiting period involves at most $n$ infections. The waiting period ends as soon as $c(I_t)\leq r/8$. Therefore, during the waiting period, infections happen at a rate of at least $r/8$. In particular, during the waiting period, the expected time between consecutive infections is at most $8/r$. For a maximum of $n$ infections, the expected time is upper bounded by $8n/r$. 
\endproof

We can now combine the various bounds we have derived so far in order to bound the expected time to extinction under our policy.

\begin{theorem} \label{thrm:upperbound}
Suppose that 
$r\geq 4W$ and that $r$ is large enough so that $np<1$, where $p$ is as defined in Lemma \ref{lem:excursionproperties}.  For any initial bag, the expected  time to extinction under the CURE policy is upper bounded by
$$\frac{1}{1-np}\cdot \frac{13n}{r}.$$
\end{theorem}
\proof{Proof:}
We start by upper bounding the expected duration of an attempt.
The length of the waiting period of an attempt is upper bounded by $8n/r$, by Lemma 
\ref{lem:detour}.

We now consider the total length of the path-following phases of an attempt. At a typical time in the path-following phase,  the current bag is on the target path, of the form $\{v_k,\ldots,v_m\}$,
where we are using the notation introduced in Section \ref{s:descr}. 
{At some point, a transition occurs and either:}
\begin{itemize}
\item[(i)]  the new bag is 
$\{v_{k+1},\ldots,v_m\}$ and the path-following phase continues, or
\item[(ii)] an infection occurs; in this case, and unless the attempt fails, the next path-following phase will again
start from the bag $\{v_{k+1},\ldots,v_m\}$. 
\end{itemize}
In both cases, the expected time spent in the path-following phase until we move to the next bag on the target path is at most $1/r$.
Since the target path has length at most $n$, 
the expected total duration of the path-following phases of an attempt is at most $n/r$.

Similarly, the number of excursions during an attempt is at most $n$. By Lemma \ref{lem:excursionproperties}, the expected length of an excursion is at most $4/r$. Therefore, the expected total time spent on excursions, during the same attempt, is upper bounded by $4n/r$.

Putting everything together, the expected duration of an attempt is at most $(8n/r)+(n/r)+(4n/r)=13n/r$.

Each attempt involves $n$ segments. During each segment, there is probability at most $p$ that the excursion is long and that the attempt fails. Therefore, the overall probability that the attempt will fail is at most $np$. Given that the process regenerates at the beginning of each attempt, the expected number of attempts is at most
$1/(1-np)$, and the desired result follows.
\endproof

\section{Corollaries and near-optimality of the CURE policy}\label{s:cor}

Theorem \ref{thrm:upperbound} has a number of interesting consequences, which we collect in the corollary that follows. 
We argue that if all nodes are initially infected, then the expected  time to extinction under any policy is  at least $n/r$. Furthermore, in a certain regime of parameters, our policy achieves $O(n/r)$ expected  time to extinction and is therefore optimal within a multiplicative constant. 
Finally, 
 if the CutWidth increases sublinearly with the number of nodes, then the expected  time to extinction can be made sublinear in $n$, using only a sublinear budget. This last result is also proved in \cite{DOT14}, using a different, nonconstructive argument.

\begin{corollary}\label{coro}
\begin{itemize}
\item[a)] For any graph with $n$ nodes and with all nodes initially infected, the expected time to extinction is at least $n/r$, under any policy.
\item[b)] 
Suppose that the budget $r$ satisfies 
$$r\geq 4W,\qquad r\geq 16 \cdot \log_2 n \cdot \Delta.$$ 
Then, for large enough $n$, and for any initial set of infected nodes, the expected time to extinction under the CURE policy is at most $26n/r$, which is sublinear in $n$.
\item[c)] Suppose that the budget $r$ satisfies the conditions in part (b), together with the condition
$$r=\Omega(n/\log n).$$ Then, the expected time to extinction under the CURE policy is of order $O(\log n)$.
\item[d)] If the CutWidth increases sublinearly with $n$, then it is possible to have sublinear time to extinction with a sublinear budget. 
\end{itemize}
\end{corollary}

\proof{Proof:}
\begin{enumerate}
\item[a)] Since nodes are cured at a rate of at most $r$, and there are $n$ nodes to be cured, the expected time to extinction must be at least $n/r$, even in the absence of infections.
\item[b)]
When $r\geq 16 \cdot \log_2 n\cdot \Delta$, 
we have $r/8\Delta \geq 2 \log_2 n$, and 
$2^{r/8\Delta} \geq n^2$. Thus,
the probability $p$ in Lemma \ref{lem:excursionproperties} is of order $O(1/n^2)$, and 
$np$ is of order $O(1/n)$. In particular, for large enough $n$, the factor $1/(1-np)$ is less than 2. By Theorem \ref{thrm:upperbound}, the expected  time to extinction is at most $26n/r$. This is sublinear in $n$, because $r$ tnds to infinity.
\item[c)] This is an immediate consequence of part (b).
\item[d)] If $W$ increase sublinearly with $n$, we can satisfy the conditions in parts (b) and (c) while keeping $r$ sublinear in $n$, and still achieve sublinear, e.g., $O(\log n)$ expected time to extinction.
\end{enumerate}
\endproof

We continue by mentioning some examples. For a line graph with $n$ nodes, the CutWidth is equal to 1 and $\Delta=2$. Therefore, by part (b) of Corollary \ref{coro} we can 
guarantee an approximately optimal expected  time to extinction, of order $O(n/r)$, as long as 
$r\geq 16 \cdot \log_2 n \cdot \Delta=32 \log_2 n$ in part (b) of Corollary. We note, however, that for this example, our analysis is not tight, and the requirement $r\geq 32 \log_2 n$ is stronger than necessary. 

For a square grid-graph with $n$ nodes, the Cut-Width is approximately $\sqrt{n}$ and $\Delta=4$.  In this case,  the requirement 
$r\geq 4W\approx 4\sqrt{n}$ is the dominant one, and suffices to  guarantee an approximately optimal expected  time to extinction, of order $O(n/r)$.  

In both of these examples, we can of course let $r$ be much larger than the minimum required, namely, $O(\log n)$ and $O(\sqrt{n})$, respectively, in order to obtain a smaller expected  time to extinction, e.g., the $O(\log n)$ expected  time to extinction in part (c) of the corollary.

\section{Conclusions}\label{concl}
{We have presented} {a} dynamic curing policy {with desirable performance characteristics. For example, if the CutWidth is sublinear in the number of nodes, the policy 
achieves sublinear  expected time to extinction, using a sublinear curing budge.} 
This policy applies to any subset of  initially infected nodes and {the resulting expected time to extinction is} order-optimal when the available budget is {sufficiently large.} 

{Our analysis brings up}  a number of open problems of   {both} practical and theoretical interest. Specifically, {a} major drawback of the CURE policy is computational complexity {because} calculating the impedance of a bag or  finding a target path {is} computationally hard. Therefore, one possible direction is the design of  computationally efficient policies with {some} performance guarantees, {perhaps for special cases.}

{Alternatively, }{In the same spirit, certain} combinatorial optimization tools have been developed for the approximation of the CutWidth of a graph. Similar tools can {perhaps} be employed to approximate the impedance of a bag. An interesting direction is the analysis of the performance of the CURE policy when instead of optimal crusades, approximately optimal crusades are used. 

{Finally, we have} argued in this paper that the CURE policy is efficient in the sense of {attaining  near-optimal, $O(n/r)$ expected time to extinction, in a certain parameter regime.
It is an interesting problem to look for approximately optimal policies over a wider set of regimes, as well as for the case where the initial set of infected nodes is small.}

\begin{APPENDIX}{Proof {of} Lemma \ref{lem:monotonewithcut}}

\def\oh{\hat\omega}

{Consider a monotone crusade $\omega\in{\mathcal{C}}(V\downarrow \emptyset)$ whose width is equal to the CutWidth $W$. This crusade starts with $V$ and removes nodes one at a time, until the empty set is obtained. Let $v_1,v_2,\ldots,v_n$ be the nodes in $V$, arranged in the order in which they are removed. }

{
Let us now fix a bag $A$. 
We construct a monotone crusade $\oh\in{\mathcal{C}}(A\downarrow \emptyset)$ as follows. We start with $A$ and remove its nodes one at a time, according to the order  prescribed by $\omega$. For example, if $n=4$, and $A=\{v_2,v_4\}$, the monotone crusade that starts from $A$ first removes node $v_2$ and then removes node $v_4$.}

{
At any intermediate step during the crusade $\oh$, the current bag  is of the form $A\cap\{v_k,\ldots,v_n\}$, for some $k$. It only remains to show that the cut of this bag is upper bounded by $\cut(A)+W$.
Let $R=\{v_1,\ldots,v_{k-1}\}$.
Note that
$$\cut(R)\leq W,$$
because of the definition of the width and the assumption that the width of $\omega$ is $W$.
Note also that the current bag is simply $A\cap R^c$.}
 
{For any two sets $S_1$ and $S_2$, let $e(S_1,S_2)$ be the number of edges that join them.
We have that 
\begin{eqnarray*}
\cut(A\cap R^c) &=& e\big(A\cap R^c, (A\cap R^c)^c\big)\\
&=&e( A\cap R^c, A^c\cup  R)\\
&\leq& e(A\cap R^c, A^c) + e(A\cap R^c, R)\\
&\leq& e(A,A^c) +e(R^c,R)\\
&=& \cut(A) +\cut(R)\\
&\leq& \cut(A)+W.
\end{eqnarray*}
We conclude that the cut associated with any intermediate bag in the crusade $\oh$ is upper bounded by $\cut(A)+W$. It follows that the width of $\oh$, and therefore $\delta(A)$ as well, is also upper bounded by that same quantity.}
\end{APPENDIX}

 \bibliography{epidemics}
 \bibliographystyle{plain}

 \end{document}